# Effects of Silica Surfaces on the Structure and Dynamics of Room Temperature Ionic Liquids: A Molecular Dynamics Simulation Study

Tamisra Pal[*], Constantin Beck, Dominik Lessnich, and Michael Vogel

Institut für Festkörperphysik

Technische Universität Darmstadt, 64289 Darmstadt, Germany

Abstract

Room temperature ionic liquids (ILs) at solid surfaces have been recognized for their significant interfacial properties in electrochemical and electronic devices. To ascertain the interface effects, we investigate dynamical and structural properties of two ILs in nanoscale confinement at various temperatures. Specifically, we perform all-atom molecular dynamics simulations for ILs composed of 1-butyl-3-methylimidazolium cations and hexafluorophosphate ([Bmim][$PF_6$]) or tetrafluoroborate ([Bmim][$BF_4$]) anions sandwiched between amorphous silica slabs. Density profiles of the ionic species across the slit reveal that [$PF_6$] and [$BF_4$] anions tend to stay closer to the slab wall than [Bmim] cations resulting in a bi-layered arrangement in the interfacial region. For the cations, we observe a preferred orientation at the surface with the methyl groups pointing towards the wall and the butyl tails projected inwards. Mean square displacements and incoherent scattering function reveal slowed and heterogeneous dynamics of all ionic species in the slit pore. In particular, spatially resolved analyses show that the structural relaxation times increase by about two orders of magnitude when approaching the silica surfaces, an effect to be considered when designing applications. The altered structural and dynamical features of the confined ILs can be related to an existence of preferred sites for the anions on the amorphous silica surfaces. Detailed analyses of relations between the broadly distributed site and surface properties show that particularly stable anion sites result when triangular arrangements of silanol groups enable multiple hydrogen bonds with the various fluorine atoms of a given anion, elucidating an important trapping mechanism at the silica surface.

---

[*] Corresponding author : Email address: tamisra.pal@physik.tu-darmstadt.de, Tel: +49 6151 16-21509

## I. Introduction

In general, solid surfaces strongly affect the properties of liquids. Due to the interaction of the liquid molecules with the surface atoms, liquid structures can be distorted and liquid dynamics can be altered. Our understanding of these effects is, however, far from complete, as a complex interplay of various energetic and geometric factors determines the behaviors. Moreover, the extent and range of structural and dynamical changes depend on temperature. A great number of studies ascertained the situation for water and polymers owing to their enormous importance in daily life.[1-4] The present work focuses on room temperature ionic liquids (ILs), which have a steadily growing spectrum of applications, leading to considerable attention in current research efforts.[5-6]

ILs are molten salts, which often consist of organic cations and inorganic anions. Due to a subtle interplay of nanoscale structural and dynamical inhomogeneity, they show complex behaviors.[5] Structural heterogeneity of ILs involves separation into polar and nonpolar domains,[7-9] which is often triggered by an amphiphilic nature of the cations. Prominent examples are ILs based on 1-alkyl-3-methylimidazolium cations with long alkyl tails. Dynamical heterogeneity, i.e., existence of transient regions with enhanced or reduced particle mobility, is typical of viscous liquids,[10-11] including ILs.[12] These effects result in complex transport properties, e.g., non-Arrhenius temperature dependence, non-exponential structural relaxation, and decoupling of translational and rotational dynamics.[12-16]

The presence of solid surfaces further complicates the structure-dynamics relations of liquids. Specifically, fluctuating interactions among the liquid molecules are complemented by a static energy landscape produced by the fixed wall atoms, leading to preferred orientations and reduced mobility. Molecular dynamics (MD) simulations proved very useful to ascertain the degree and range of interface effects.[17-24] Since full microscopic information is available, this method enables investigations with unparalleled spatial and temporal resolution. These possibilities yielded valuable insights into water and polymer behaviors at a large number of different interfaces. [5, 17-24]

Relating to ILs, experimental and computational approaches observed various ordering phenomena at solid surfaces,[25-29] which are of potential relevance for rapidly growing application fields.[5, 30-33] In particular, the formation of density layers and existence of preferred orientations lead to crystalline-like structures in the interfacial region. Pioneering experimental work focused on 1-alkyl-3-methylimidazolium-based systems on silica substrates.[34] The results showed that the methyl groups points towards the silica surface, whereas the alkyl tails point towards the liquid reservoir. Further experimental studies reported similar ordering phenomena for ILs on various types of amorphous or crystalline substrates.[25, 29, 35-38] However, the degree and type of order strongly depend on the nature of the solid surface and on the properties of the cations and, in particular, the anions. MD simulation work systematically varied the size and shape of the cations and anions.[39] Detailed calculation of interaction energies with α-Quartz revealed that anion contributions dominate. Computational studies on the role of the surface structure found that ordered interfacial layers are less prominent at amorphous silica than at crystalline silica, explicitly, β-cristobalite,[40] or in carbon nanotubes.[41] Moreover, charged templates were used to explore surface ordering under conditions relevant for applications.[42]

While such studies provided manifold insights into changes of IL structures, work on IL dynamics near solid surfaces is still sparse. Measurements and simulations revealed that the presence of silica walls leads to a strong slowdown of cation and anion dynamics.[29, 43] For a detailed characterization, MD simulation studies exploited the possibility to perform spatially resolved analyses. In such approaches, prominent mobility gradients were observed across silica and graphitic pores.[41, 44]

Here, we perform MD simulations to investigate the structure and, in particular, the dynamics of IL models in a slit-like pore formed by amorphous silica slabs. In this way, we intend to improve our understanding of structure-dynamics relations under confinement conditions. Considering that the properties of the anion may play a crucial role, we pair the much studied 1-alkyl-3-methylimidazolium [Bmim] cations with hexafluorophosphate [PF$_6$] or tetrafluoroborate [BF$_4$] anions, resulting in hydrophobic or hydrophilic ILs, respectively.[45-47] We show that detailed insights into the degree and range of surface effects are available when we determine structural and dynamical properties as a function of the distance from the silica wall. Furthermore, we demonstrate that valuable knowledge about the mechanisms underlying the altered behaviors can be obtained when we identify the adsorption sites at the amorphous surface directly from the ionic trajectories.

## II. Methodology

### A. Simulation Details

We use GROMACS version 5.1.3[48] for performing all-atom simulations of [Bmim][PF$_6$] and [Bmim][BF$_4$] in a silica pore. Fig. 1 shows the chemical structure of the cation and the respective anions. To match the [Bmim][PF$_6$] and [Bmim][BF$_4$] bulk density of 1.369 g/cm$^3$ and 1.20 g/cm$^3$ at room temperature, respectively,[49-50] we use 367 ion pairs and adjust the pore volume accordingly, see below. Depending on temperature, the resulting sizes of the simulation boxes amount to about (4.03 x 4.14 x 9.62) nm$^3$ for [Bmim][PF$_6$] and (4.03 x 4.15 x 9.15) nm$^3$ for [Bmim][BF$_4$].

The bonded interactions of the ions include terms for bonds and angles and, in the case of the cations, for dihedrals. A force field developed by Balasubramanian et al. is employed for the explicit atom model of [Bmim][PF$_6$].[51-52] It uses charges of 0.8$e$ and -0.8$e$ for the cations and anions, respectively. For [Bmim][BF$_4$], the interaction parameters of [Bmim] are the same, while that of [BF$_4$] are taken from Padua and Lopes.[53] The parameters for the non-bonded interactions are taken from Sambasivarao et al.[54] and the atomic masses are taken from Weiser et al.[55] The partial charges of the atoms in [BF$_4$] are uniformly scaled to obtain a total charge of -0.8$e$. This concept of charge scaling to +/-0.8$e$ has been adopted to arrive at a better description of transport properties. Altogether, the potential form is based on the OPLS-AA/AMBER framework and identical with that used by the Lopes et al.[52]

$$V(\{\mathbf{r}\}) = \sum_{ij}^{bonds} \frac{k_{r,ij}}{2}(r_{ij} - r_{0,ij})^2 + \sum_{ijk}^{angles} \frac{k_{\theta,ijk}}{2}(\theta_{ijk} - \theta_{0,ijk})^2 + \sum_{ijkl}^{dihedrals} \sum_{m=1}^{6} V_{m,ijkl} \cos^m(\phi_{ijkl})$$
$$+ \frac{1}{2} \sum_i \sum_{j \neq i} \left\{ 4\varepsilon_{ij} \left[ \left(\frac{\sigma_{ij}}{r_{ij}}\right)^{12} - \left(\frac{\sigma_{ij}}{r_{ij}}\right)^{6} \right] + \frac{q_i q_j}{4\pi\varepsilon_0 r_{ij}} \right\}$$
(1)

Preliminary IL configurations are generated using Avogadro 1.1.1[56] and Packmol,[57] followed by an energy minimization employing the steepest descent algorithm.

In all simulations, the amorphous silica slab has a volume of (4.03 x 4.14 x 2.41) nm$^3$. The force field parameters as well as the initial atomic positions are taken from the work of Emami et al.[58] The silica slab is placed parallel to the *xy* plane on one side of the simulation box so that a slit pore results from application of periodic boundary conditions in three dimensions. We fix Si and O atoms in the middle of the silica slab using the position restraints algorithm to obtain a time-independent pore geometry. The silica slab has a chemically realistic surface, in particular, it features silanol groups with a surface density of 4.7 silanol groups /nm$^2$. The hydroxyl groups of these entities have rotational mobility about the Si-O bonds. The non-bonded interactions at the IL-silica interface are determined by the Lorentz-Berthelot mixing rules.

To integrate the equations of motion, we use the leapfrog algorithm [59] with a time step of 1 fs. The electrostatic interactions are computed employing the Particle-Mesh-Ewald method.[60-61] The electrostatic cut-off in real space is set to 14 Å for [Bmim][PF$_6$] and 12 Å for [Bmim][BF$_4$]. The Fourier spacing in reciprocal space is 0.15 nm$^{-1}$. Equilibrated structures are obtained from NPT simulations. First, we perform NPT runs utilizing the Berendsen thermostat and barostat [62] for temperature and pressure couplings for about 20 ns at 500 K. Afterwards, we employ NPT runs with the Nose-Hoover thermostat[63-64] and the Parrinello-Rahman barostat[65] for stepwise temperature annealing. The annealing starts at 500 K and involves temperature steps of 50 K, which are applied every 20 ns until the desired temperature is reached. The system configurations obtained from this equilibration process are used as initial frames of the NVT production runs, which utilize the Nose-Hoover thermostat with a time constant of 1.3 ps and separate temperature coupling for cations and anions. The total length of the production runs amounts to 200 ns for higher temperatures and 400 ns for lower temperatures.

### B. Analysis Protocol

To analyze surface effects, we first study density profiles perpendicular to the silica surface, i.e., in $z$ direction. Specifically, we determine the time-averaged number density $n_z$ for the cations and anions using their center-of-mass positions and bins of a width of 0.024 nm and normalize the results by the bulk number density $n_0$. As this analysis reveals pronounced density layers of a thickness of ca. 0.5 nm, see below, we separately determine structural and dynamical properties in these layers. For this purpose, we group the ions into layers of 0.5 nm thickness based on the distance between their center of mass and the nearest atom of the silica wall at a given time. The latter criterion allows us to consider the surface roughness. The layers are numbered sequentially from the nearest to the farthest from the silica wall. In studies of preferential orientation, we focus on the cations in the first layer and calculate the angles between vectors suitable to describe the direction of specific molecular entities and the $z$ axis. We use the vector connecting the $N_A(1)$ and $C_T$ atoms, see Fig. 1, to characterize the orientation of the butyl tail and the surface normal of the imidazolium plane to specify the orientation of the ring entity. A schematic representation of these vectors is given in the corresponding figures. In spatially resolved analyses of dynamical properties, we discriminate ions in different layers based on their center-of-mass positions at the beginning of a studied time interval, $t_0$.

For spotting adsorption sites of $[PF_6]$ and $[BF_4]$ at the silica surface, we determine areas of high anion densities in the first layer directly from the simulated configurations. Specifically, we use the Gaussian kernel density estimation technique to gauge the probability densities of finding the P and B atoms at $(x,y,z)$ coordinate points in the first layer based on all atomic positions in this region during the simulation run. The obtained probability distributions show that highly populated areas are well separated by regions with nearly vanishing occupation probability. Hence, it is possible to identify favored areas with adsorption sites of the anions. In detail, we define sites as connected spatial regions in which the probability density of finding an anion amounts to at least the average value. The center-of-mass positions $\vec{r}_{cs}$ for the sites $s$ can be defined as

$$\vec{r}_{cs} = \frac{\sum_{f}^{N_f} \sum_{s}^{N_s(f)} \vec{r}_i^s(f)}{\sum_{f}^{N_f} N_s(f)} \qquad (2)$$

The sum involves the anion coordinates ($\vec{r}_i^s$) of the $N_s$ anions occupying site $s$ at a given time frame $f$, and the time average over all $N_f$ time frames. Based on this, we determine a characteristic radius $r_{V_s}$ and the volume $V_s$ of each site using the convex-hull method to approximate the shape of polygonal sites for volume calculation. Moreover, we define a limiting radius $r_{\lim}$ to cut-off the distance beyond which the site is not considered to be occupied by any anion. This value is set to $r_{\lim} = 2r_{V_s}$. Altogether, we define that site $s$ is occupied by an anion at a given time frame $f$ if $\left| \vec{r}_{cs} - \vec{r}_i^s(f) \right| < r_{\lim}$.

For unravelling the origin and nature of these adsorption sites, it is important to identify silanol groups that can form hydrogen bonds with an anion residing at a given site. We determine hydrogen bonds between [PF$_6$] or [BF$_4$] and the silica surface using the position vectors of the fluorine atoms of the anions ($\vec{r}_F$) and of the oxygen and hydrogen atoms of the silanol groups ($\vec{r}_O$ and $\vec{r}_H$). Specifically, adopting the usual geometric criterion for hydrogen bonds in aqueous systems,[66] we request that all of the following three criteria need to be fulfilled in order to consider an anion-silanol configuration as a hydrogen bond:

$$\begin{aligned} \left| \vec{r}_F - \vec{r}_H \right| &< 0.267 nm \\ \left| \vec{r}_F - \vec{r}_O \right| &< 0.3615 nm \\ \theta_{F-O-H} &< 30^o \end{aligned} \qquad (3)$$

Here, $\theta_{F-O-H}$ is the angle between the directions of the O-H bond and the O-F vector. The first criterion is obtained from the van der Waals radii of the F and H atoms in the force field, which amount to 0.147 and 0.120 nm, respectively. The second one derives from the first one in combination with the O-H bond length of 0.0945 nm in the interaction potential. Finally, the third criterion considers the directionality of hydrogen bonds, in analogy with the widely used definition of hydrogen bonds in water.[66] Similar criteria were applied in previous X-ray studies[67-68] on organo fluorine compounds to study halogen-hydrogen bonds. If, based on these criteria, a silanol group can form a hydrogen bond with an anion residing at any position within a given adsorption site, it is assigned to this site. The number of silanol groups belonging to the various adsorption sites are denoted as $n_{SiOH}$ in the following.

## III. Results

### A. Structural Properties

First, we focus on the number density profiles of the cations and anions across the slit. Fig. 2 shows results for [Bmim][PF$_6$] at 300K. While there are pronounced layers near the silica walls, density oscillations are weak in the slit center. To further investigate the layering, the inset focuses on the interfacial region and compares findings for [Bmim][PF$_6$] and [Bmim][BF$_4$]. The number density profiles reveal that the layering effect is stronger for the anions than for the cations and that the former are predominantly closer to the silica wall than the latter. These findings are reasonable as [PF$_6$] and [BF$_4$] anions are more spherically symmetric and smaller than [Bmim] cations. Likewise, [BF$_4$] is located somewhat closer to the silica surface than [PF$_6$], because the former has a smaller ionic radius of 4.58 Å than the latter with 5.44 Å. A formation of density layers is corroborated by results from atomic force microscopy measurements,[29] which suggest a first layer with crystal-like ordering followed by alternating bilayers of cations and anions in the interfacial region. Furthermore, our observations match reflectivity experiments,[69] revealing enhanced electron density for [Bmim][BF$_4$] and [Bmim][PF$_6$] near a surface. The thickness of the layers is also consistent with experimental estimates in the range 0.4-0.8 nm,[29, 36, 42, 69] depending on the length of the alkyl chain attached to the imidazolium ring of the cation.

Next, we analyze the orientation of the [Bmim] cations in the first layer at the silica surface. First information is available from number density profiles of different types of atoms (not shown). The preferential positions of carbon $C_1$ in the methyl group and of carbon $C_R$ in the ring are closer to the surface than that of the terminal carbon $C_T$ of the butyl tail, suggesting that the methyl group points towards the surface and the alkyl tail away from it. Further insights into the interaction of the [Bmim] cations with the silica surface can be obtained from angles $\theta$ characterizing the orientations of specific molecular groups with respect to the normal vector of the silica surface, i.e., with the $z$ axis of the simulation box. Specifically, $\theta_{butyl}$ specifies the orientation of the end-to-end vector of the butyl chain and $\theta_{ring}$ that of the vector normal to the ring plane. In Fig. 3, we present the probability distribution of $\cos\theta_{butyl}$ for [Bmim][PF$_6$] and [Bmim][BF$_4$] at 300 K. We see two maxima corresponding to two preferred arrangements of the butyl tail: (i) a tilted orientation with angles $\theta_{butyl} \approx$ 45-60° ($\cos\theta_{butyl} \approx$ 0.5-0.7) and (ii) an orientation with angles $\theta_{butyl} \approx 105°$ ($\cos\theta_{butyl} \approx$ -0.25), i.e., nearly parallel to the silica surface. For both ILs, the tilted conformation dominates. However, the peaks are sharper and, hence, the configurations are better defined for [Bmim][PF$_6$] than for [Bmim][BF$_4$], suggesting that [PF$_6$] involves a more specific orientation of the [Bmim] cation than [BF$_4$]. These results are again in harmony with experimental data. In detail, Wipff et al.[39] observed butyl orientations at ~60° and ~116° for the [Bmim] cations on α-Quartz surface. Moreover, Conboy and coworkers[34] found that an increase in the size of the anion enhances orientations of alkyl tails perpendicular to the surface normal. Figure 4 shows the probability distribution of $\cos\theta_{ring}$ at 300 K. We see that angles $\theta_{ring}$ of about 0° and 180° are favored, indicating that the imidazolium rings tend to be aligned parallel to the silica surface. Statements about a dependence on the type of the anion are inconclusive due to limited statistics. Nevertheless, our simulation study complements findings of Wipff et al.,[39] who reported $\theta_{ring} \approx 40°$ at α-Quartz surface. However, one should bear in mind that the silica surface is amorphous and rugged in our case. To illustrate typical cation orientations at the silica surface for [Bmim][PF$_6$] and [Bmim][BF$_4$], the cations residing in the first layer at a given time frame are shown together with the surface contour in Fig. 5. We see that a considerable roughness of the silica surface adds to significant distributions for the configurations and orientations of the cations, as quantified in Figs. 3 and 4.

B.   **Dynamical Properties**

For an investigation of ion dynamics, we first determine the mean square displacement (MSD)

$$r^2(t) = \left\langle \left[ \vec{r}_i(t+t_0) - \vec{r}_i(t_0) \right]^2 \right\rangle \qquad (4)$$

It relates the center-of-mass position of an ion, $\vec{r}_i$, at two times separated by a time interval $t$. The pointed brackets denote averages over all cations or anions and various time origins $t_0$. Figure 6 shows MSD data for [Bmim][PF$_6$] and [Bmim][BF$_4$] in silica confinement at 300 K. Typical of viscous liquids, three regions can be distinguished. The initial $r^2(t) \propto t^2$ behavior accounting for ballistic motion is followed by a sub-diffusive regime associated with a temporary trapping of the ions by their neighbors, and finally a linear regime $r^2(t) \propto t$ indicative of diffusive motion. Figs. 6(a) and 6(b) display the MSD of the cations and anions, respectively. Both ionic species diffuse faster in [Bmim][BF$_4$] than in [Bmim][PF$_6$], in agreement with results of a pulsed field gradient NMR study.[49]

For quantitative analysis, we determine self-diffusion coefficients $D$ from the linear regime using the relation $r^2(t) = 6Dt$. In doing so, we exploit that the covered time and length scales are relatively small so that a crossover from 3D to 2D diffusion expected for a slit geometry has only minor effects. In Fig. 7, we compile temperature-dependent diffusivities of the ionic species in bulk and in confinement. For cations and anions, the mobility is higher in the bulk than in the confinement over a wide range of temperatures. In all cases, the anions exhibit slower diffusion than the cations. This behavior is typical of [Bmim] based ILs.[70] The temperature dependence is weaker for [Bmim][BF$_4$] than for [Bmim][PF$_6$] both in bulk and in confinement. In detail, the diffusivities differ by roughly a factor of two between both bulk samples at 400 K, while this difference increases to about an order of magnitude at 260 K. For the bulk ILs, clear deviations from Arrhenius behavior are observed, which can be described by the Vogel-Fulcher-Tammann (VFT) equation, $\tau = \tau_0 \exp\left( \frac{E}{k_B(T - T_\infty)} \right)$. In the silica pore, the non-Arrhenius behavior is less prominent, in particular, for [Bmim][BF$_4$]. However, the limited temperature range and the available data quality do not allow us to arrive at definite conclusions about the latter aspect.

To investigate the effect of silica surfaces on cation and anion dynamics in more detail, we preform spatially resolved analyses. Specifically, we separately calculate incoherent scattering functions

$$S(q,t) = \left\langle \cos\left[\vec{q}\left(\vec{r}_i(t+t_0) - \vec{r}_i(t_0)\right)\right]\right\rangle \tag{5}$$

for ions in different density layers at the surface. The modulus of the scattering vector $q = |\vec{q}|$ is set to values corresponding to the nearest neighbor distance of the cation and anion, which is 14.50 nm$^{-1}$ for [Bmim][PF$_6$] and 13.96 nm$^{-1}$ for [Bmim][BF$_4$]. We distinguish four layers of 0.5 nm thickness based on the distances of the center-of-mass positions of the ions to the surface at the respective time origin $t_0$. Typical of viscous liquids, $S(q,t)$ decays in two steps separated by an extended plateau regime resulting from a temporary trapping of the ions by their neighbors. Here, we focus on the long-time decay due to structural ($\alpha$) relaxation. Considering the stretched exponential form of this decay, we determine the structural relaxation time $\tau_\alpha$ in a density layer by fitting to the Kohlrausch-Williams-Watts (KWW) function

$$S(q,t) \approx A \exp\left(-\left(\frac{t}{\tau_\alpha}\right)^\beta\right) \tag{6}$$

where the stretching exponent $\beta$ characterizes the degree of the non-exponentiality.

Fig. 8 shows the spatially resolved scattering functions for [Bmim][PF$_6$] at 300 K. For both cations and anions, we see that the decays of $S(q,t)$ strongly shift to longer times when approaching the silica wall, indicating a pronounced slowdown of ion dynamics. The [Bmim] dynamics is almost two orders of magnitude slower in the first layer at the surface than in the fourth layer, where bulk behavior is nearly recovered. The difference between the structural relaxation in various pore regions is even larger for [PF$_6$] dynamics. Thus, there is an enormous gradient of the ionic mobilities across the slit. The observation of a stronger slowdown of anion dynamics as compared to cation dynamics is consistent with the finding that the preference to reside near the silica wall is higher for the former than the latter ionic species.

In Fig. 9, we explore the anion dynamics in the slit pore for [Bmim][PF$_6$] and [Bmim][BF$_4$] based on the temperature-dependent time constants $\tau_\alpha$. First, we focus on results obtained from an average over the whole pore volume. Consistent with the findings for the diffusion coefficients $D$, the pore-averaged structural relaxation times indicate that [PF$_6$] is slower than [BF$_4$]. The ratio $\tau_\alpha^{PF_6}/\tau_\alpha^{BF_4}$ increases with decreasing temperature and amounts to about an order of magnitude at 300 K. For both anions, the temperature dependence of $\tau_\alpha$ in silica confinement deviates, if at all, only weakly from an Arrhenius law. Comparison of the time constants $\tau_\alpha$ for the different layers provides access to the effect of the silica walls on anion dynamics. We see that anion dynamics is drastically slower in the first layer than in the third layer. This discrepancy becomes more prominent upon cooling and it is stronger for [PF$_6$] than for [BF$_4$]. Consistently, the above analyses revealed more prominent density layers and preferred orientations for the former than the latter anions.

In studies on confined water,[17-18] it was argued that a comparable slowdown of molecular dynamics at solid interfaces can be attributed to the fact that the fixed wall atoms produce a static energy landscape, e.g. by providing sites for hydrogen bonding, to which the neighboring liquid has to adapt, limiting the possibility for structural rearrangements in the interfacial region. In our case, the silanol groups on the silica surface also furnish hydrophilic sites to which the anions can be anchored via transient weak hydrogen bonds. In the following, we, therefore, investigate the role of anion sites at the silica wall to obtain closer insights into the mechanisms underlying altered IL behaviors at solid interfaces.

### C.    Characterization of Adsorption Sites

Areas on the silica surface with a high density of silanol groups may promote hydrogen bond formation. However, hydrogen bonds involving halogen atoms are relatively weak with energies comparable to that of van der Waals complexes. Nevertheless, X-ray studies[67-68, 71] and theoretical calculations[67-68] reinforced the bonding capabilities of halogen atoms with hydroxyl hydrogens. Due to a high electronegativity, the fluorine atoms of the anions act as hydrogen bond acceptors[72] for the hydrogen atoms of the silanol groups. As a consequence of the weakness of these interactions, we do not determine adsorption sites at the silica surface from energetic criteria, but rather from the ionic trajectories, see Sec. II.B. A similar approach was taken to determine ion sites in glassy electrolytes.[73-74] Specifically, we exploit that the probability densities for the anionic positions at the silica surface show regions with high occupancies, which are separated by depopulated areas. Thus, the former regions can be identified with preferred sites of the anions.

In the following, we investigate the distribution of [PF$_6$] and [BF$_4$] anions on the silica surface at 350 K. This temperature is a compromise between anion dynamics becoming too slow for site exchange at low temperatures and thermal energies becoming too high for site identification at high temperatures. 2D plots of regions in the first layer with high probability of finding [PF$_6$] and [BF$_4$] anions are shown in Figs. 10(a) and 10(b), respectively. Various preferred adsorption sites are clearly distinguishable. For clarity, the different sites are depicted in different colors and labelled by consecutive numbers. Also, the respective centers of the sites, $\vec{r}_{cs}$, and the positions of the silanol groups are marked. The total number of sites obtained from 200 ns simulation runs is 17 for [PF$_6$] and 24 for [BF$_4$]. Thus, there are fewer [PF$_6$] sites than [BF$_4$] sites, most probably due to the diverse size and geometry of the anion species. As expected for an amorphous surface, the positions of the sites are not regularly distributed. Moreover, the sites have different shapes, in particular, triangular shapes are striking for [PF$_6$]. For most sites, the extension in $z$ direction is small, warranting a 2D representation.

In Fig. 11, we show characteristic properties of the identified sites, explicitly, the volume $V_s$ of the spatial region attributed to the site, the time-averaged number $\langle n_{anion} \rangle$ of anions occupying the site, and the number $n_{SiOH}$ of silanol hydroxy groups assigned to the site. More detailed definitions of these quantities were given in Sec. II.B. In Fig. 11(a), we observe that, on average, the sites of the smaller [BF$_4$] anions are almost twice as big as that of the larger [PF$_6$] anions. On first glance, this result may be surprising. However, it is necessary to consider that the sites are obtained from the preferred positions of the central B and P atoms and, hence, their definition does not involve the anionic radii. Rather, broader regions of enhanced probability density for [BF$_4$] than [PF$_6$] imply that the corresponding energy minima are shallower for the former than the latter anion species. Most probably, this difference in the local energy landscape is at the origin of our finding that the slowdown due to the silica surface is weaker for [BF$_4$] than for [PF$_6$]. In Fig. 11(b), it is evident that the average occupancy of [BF$_4$] and [PF$_6$] sites is comparable. However, for both anion species, $\langle n_{anion} \rangle$ strongly varies from site to site. While there is, if at all, only a weak correlation between the occupancy and the size of a site, more and less frequently populated sites differ with respect to their shapes. Specifically, for [PF$_6$], sites with numbers 1-3, 7, and 10-12 have the highest occupancies. Interestingly, all these sites have triangular shape, as can be inferred from closer inspection of Fig. 10(a). Moreover, they often have reverse triangular arrangements of silanol groups in their neighborhoods. By contrast, these highly populated sites are not conspicuous with respect to the number $n_{SiOH}$ of silanol groups available for hydrogen bond formation when located at these spots. Analogous relations hold for the [BF$_4$] sites. These findings imply that good absorption sites for the anions are not distinguished by a large size or a high density of silanol groups, but rather by a suitable, most probably, triangular arrangement of silanol groups so as to enable simultaneous hydrogen-bond formation. Consistently, Canova et al.[40] reported evidence for triangular geometries for ILs on a crystalline silica surface.

### D.     Nature of Hydrogen Bonds

Finally, we ascertain the nature and role of the weak hydrogen bonds between the [PF$_6$] and [BF$_4$] anions and the silanol groups of the silica surface. By using the hydrogen-bond criteria given in Sec II.B, we determine the time-averaged number of hydrogen bonds formed by each of the silanol groups with the given anion species, $\langle n_{HB} \rangle$. In Fig. 12, we show the cumulative distribution of this quantity, $C(\langle n_{HB} \rangle)$, for [PF$_6$] and [BF$_4$] anions at 350 K. For [PF$_6$], we see that about 40% of the silanol groups are not involved in hydrogen bonds and another ~25% are characterized by values $\langle n_{HB} \rangle$ below 0.3, followed by a plateau of the cumulative distribution up to $\langle n_{HB} \rangle \approx 0.5$. Thus, ca. 2/3 of the silanol groups have little importance for the interaction with [PF$_6$], while the remaining 1/3 of the groups is responsible for the strong adsorption of these anions at the silica surface. For [BF$_4$], the situation is inverse. Fractions of about 1/3 and 2/3 are rarely and frequently involved in hydrogen bonds with these anions. These findings suggest that [PF$_6$] has more stringent requirements regarding the spatial arrangement of the hydrogen-bond donors so that a smaller fraction of silanol groups meets the demands for these anions than for [BF$_4$]. On the other hand, when suitable, often triangular configurations of silanol groups are available, [PF$_6$] is capable of forming relatively stable bonds with the silica surface, as indicated by a prominent slowdown of dynamics in the interfacial region.

Interestingly about 20% of the silanol groups show $\langle n_{HB} \rangle > 1$ for both anion types. In particular, the distribution well extends up to $\langle n_{HB} \rangle = 1.2$ for [PF$_6$]. This means that substantial fractions of silanol groups are involved in more than one hydrogen bond at a time, implying that they simultaneously interact with two of the fluorine atoms of the [PF$_6$] and [BF$_4$] anions. To consider such situations, the concept of furcated bonds can be introduced. In the literature, this concept was applied to disordered states of water.[75] Moreover, bi-furcated bonds were reported for a bulk IL.[76] In Fig. 13, we picturize exemplary hydrogen-bond configurations for [PF$_6$] and [BF$_4$] anions at an adsorption site. For both types of anions, non-furcated and furcated hydrogen bonds coexist. Moreover, we see for [PF$_6$] that, due to these diverse bonding geometries, five of the six fluorine atoms can be involved in hydrogen bonds at the same time. One can expect that such high number of hydrogen bonds enables particularly stable configurations and, hence, leads to a strong slowdown of [PF$_6$] dynamics at these sites.

## IV.    Conclusion

We have discussed results from MD simulations of two room temperature ILs, [Bmim][PF$_6$] and [Bmim][BF$_4$], confined in a ca. 6 nm wide slit between amorphous silica slabs. We have found that both structural and dynamical properties of these ILs are strongly affected by the silica surfaces. Although ILs are mainly governed by electrostatic interactions, we have observed that hydrogen bonds between the [PF$_6$] or [BF$_4$] anions and the silanol groups of the silica walls play a prominent role for the surface effects. To obtain detailed insights, it has been exploited that spatially resolved analyses provide access to the degree and range of structural and dynamical distortions in the interfacial region over a wide temperature range and that an identification of adsorption sites from ionic trajectories yields valuable information about the underlying mechanisms.

We have found that several properties obtained from our simulation approach well agree with experimental observations,[34, 36, 69] confirming the high fidelity of the used [Bmim][PF$_6$] and [Bmim][BF$_4$] models. Number density profiles have revealed prominent density layering near the interfaces, consistent with previous results for various ILs at solid surfaces.[34, 40, 42, 69, 77] Therefore, our analyses of IL structure and dynamics have separately ascertained the behaviors in different layers. Relating to structure, we have observed that the [Bmim] cations show preferred orientations in the first layer. The methyl group and butyl tail prefer to point towards the silica wall and the pore center, respectively, and the imidazolium ring tends to align with the silica surface. However, the orientation of the cations in the first layer depends on the type of the anions, which reside closer to the silica surface depending on their molecular shape and size. As for dynamics, we have found that [Bmim][PF$_6$] and [Bmim][BF$_4$], overall, show significantly slower self diffusion and structural relaxation in the pore than in the bulk. Moreover, deviations from Arrhenius behavior are weaker for the confined ILs than for the bulk ILs, which show the characteristic fragile behavior of molecular glass formers. Our spatially resolved analysis enabled deeper insights into the origins. While the dynamics in the center of the slit resembles that in the bulk, cation and, in particular, anion motions are slowed down by several orders of magnitude at the surfaces. This effect increases with decreasing temperature and decreases with increasing distance from the surface. Thus, strong mobility gradients exist across the slit pore, which become more prominent upon cooling and involve about four surface layers in the studied temperature range. Interestingly, the slowdown of ion dynamics at the silica surface is more prominent for [Bmim][PF$_6$] than for [Bmim][BF$_4$]. This result is unexpected since the latter IL is considered more hydrophilic than the former IL and, hence, one might have argued that [Bmim][PF$_6$] has stronger interactions with the hydrophilic silica surface than [Bmim][BF$_4$].

To elucidate the molecular mechanisms underlying these surface effects, detailed studies of anion-silica interactions have proven useful. Careful analysis of the simulated trajectories has revealed that there are well defined adsorption sites for the anions at the silica surface. Due to the amorphous nature of the walls, the adsorption sites show, however, broad distributions of properties. We have found that [PF$_6$] sites have, on average, a smaller volume than [BF$_4$] sites, implying sharper and deeper local energy minima for the former than the latter anion species. We propose that this effect is at the origin of our finding that the slowdown at the silica surfaces is more prominent for [PF$_6$] than [BF$_4$]. The suitability of the sites to host an anion, as measured, by the occupancy, is not governed by the sheer number of silanol groups available for hydrogen bonding, but rather by their spatial arrangement on the surface. Specifically, sites with high occupancies often have triangular shapes, which reflect an inverse triangular arrangement of neighboring silanol groups. Studying the weak hydrogen bonds between the fluorine atoms of the anions and the silica wall in some detail, it has been shown that large fractions of the silanol groups on the surface, for [PF$_6$] a fraction of ~2/3, are hardly involved in hydrogen bonding. This effect can be explained by the fact that hydrogen bonds between fluorine atoms and hydroxy groups are relatively weak so that several of these bonds need to be formed for strong anion-silica interaction and, hence, a suitably arranged set of hydrogen-bond donors rather than a single silanol group are required to create a good adsorption site. Accordingly, we have found that anions in adsorption sites form several hydrogen bonds, including furcated bonds. For [PF$_6$], simultaneous formation of up to five hydrogen bonds has been observed. Thus, our results reveal that the particularly prominent slowdown of [PF$_6$] at the silica surface can be rationalized based on the anion and surface structures.

In conclusion, the properties of ILs are significantly altered near solid surfaces. The nature of these changes differs for various ILs and depends on the type of the surface. The present simulation approach has helped to improve our microscopic understanding. In particular, it has unraveled that the surface effects are strongly governed by the structure and shape of the ions and the spatial distribution of surface groups available for binding, e.g., for hydrogen-bond formation. These altered behaviors need to be considered when designing new applications involving interfaces such as electrodes or supports in film technologies. In doing so, the significant dependence on the ion and surface structures, in particular, their geometries, can provide guiding principles to tailor the IL properties. Thus, our fundamental approach and theoretical understanding relates to the rapidly developing field of high performance IL applications.

**Acknowledgement**

This research has been supported by the Deutsche Forschungsgemeinschaft (DFG) through the Collaborative Research Center/Transregio TRR 146 (Project A6).

**References**

1. Stanley, H. E.; Buldyrev, S. V.; Franzese, G.; Kumar, P.; Mallamace, F.; Mazza, M. G.; Stokely, K.; Xu, L., Liquid Polymorphism: Water in Nanoconfined and Biological Environments. *J. Phys. Condens. Matter* **2010**, *22*, 284101.
2. Cerveny, S.; Mallamace, F.; Swenson, J.; Vogel, M.; Xu, L., Confined Water as Model of Supercooled Water. *Chem. Rev.* **2016**, *116*, 7608-7625.
3. Giovambattista, N.; Rossky, P. J.; Debenedetti, P. G., Computational Studies of Pressure, Temperature, and Surface Effects on the Structure and Thermodynamics of Confined Water. *Annu. Rev. Phys. Chem.* **2012**, *63*, 179-200.
4. McKenna, G. B., Ten (or More) Years of Dynamics in Confinement: Perspectives for 2010. *Eur. Phys. J. Special Topics* **2010**, *189*, 285-302.
5. Hayes, R.; Warr, G. G.; Atkin, R., At the Interface: Solvation and Designing Ionic Liquids. *Phys. Chem. Chem. Phys.* **2010**, *12*, 1709-1723.
6. Hayes, R.; Warr, G. G.; Atkin, R., Structure and Nanostructure in Ionic Liquids. *Chem. Rev.* **2015**, *115*, 6357-6426.
7. Del Pópolo, M. G.; Voth, G. A., On the Structure and Dynamics of Ionic Liquids. *J. Phys. Chem. B* **2004**, *108*, 1744-1752.
8. Canongia Lopes, J. N.; Costa Gomes, M. F.; Pádua, A. A. H., Nonpolar, Polar, and Associating Solutes in Ionic Liquids. *J. Phys. Chem. B* **2006**, *110*, 16816-16818.
9. Triolo, A.; Russina, O.; Bleif, H.-J.; Di Cola, E., Nanoscale Segregation in Room Temperature Ionic Liquids. *J. Phys. Chem. B* **2007**, *111*, 4641-4644.
10. Ediger, M. D., Spatially Heterogeneous Dynamics in Supercooled Liquids. *Annu. Rev. Phys. Chem.* **2000**, *51*, 99-128.

11. Berthier, L.; Biroli, G., Theoretical Perspective on the Glass Transition and Amorphous Materials. *Rev. Mod. Phys.* **2011**, *83*, 587-645.
12. Weingärtner, H., Nmr Studies of Ionic Liquids: Structure and Dynamics. *Curr. Opin. Colloid Interface Sci.* **2013**, *18*, 183-189.
13. Cicerone, M. T.; Ediger, M. D., Enhanced Translation of Probe Molecules in Supercooled O-Terphenyl: Signature of Spatially Heterogeneous Dynamics? *J. Chem. Phys.* **1996**, *104*, 7210-7218.
14. Funston, A. M.; Fadeeva, T. A.; Wishart, J. F.; Castner, E. W., Fluorescence Probing of Temperature-Dependent Dynamics and Friction in Ionic Liquid Local Environments. *J. Phys. Chem. B* **2007**, *111*, 4963-4977.
15. Zhang, X.-X.; Liang, M.; Ernsting, N. P.; Maroncelli, M., Complete Solvation Response of Coumarin 153 in Ionic Liquids. *J. Phys. Chem. B* **2013**, *117*, 4291-4304.
16. Sengupta, S.; Karmakar, S.; Dasgupta, C.; Sastry, S., Breakdown of the Stokes-Einstein Relation in Two, Three, and Four Dimensions. *J. Chem. Phys.* **2013**, *138*, 12A548.
17. Klameth, F.; Vogel, M., Structure and Dynamics of Supercooled Water in Neutral Confinements. *J. Chem. Phys.* **2013**, *138*, 134503.
18. Harrach, M. F.; Klameth, F.; Drossel, B.; Vogel, M., Effect of the Hydroaffinity and Topology of Pore Walls on the Structure and Dynamics of Confined Water. *J. Chem. Phys.* **2015**, *142*, 034703.
19. Scheidler, P.; Kob, W.; Binder, K., The Relaxation Dynamics of a Simple Glass Former Confined in a Pore. *EPL* **2000**, *52*, 277.
20. Varnik, F.; Baschnagel, J.; Binder, K.; Mareschal, M., Confinement Effects on the Slow Dynamics of a Supercooled Polymer Melt: Rouse Modes and the Incoherent Scattering Function. *Eur. Phys. J. E* **2003**, *12*, 167-171.
21. Vogel, M., Rotational and Conformational Dynamics of a Model Polymer Melt at Solid Interfaces. *Macromolecules* **2009**, *42*, 9498-9505.
22. Mirigian, S.; Schweizer, K. S., Communication: Slow Relaxation, Spatial Mobility Gradients, and Vitrification in Confined Films. *J. Chem. Phys.* **2014**, *141*, 161103.
23. Lerbret, A.; Lelong, G.; Mason, P. E.; Saboungi, M.-L.; Brady, J. W., Water Confined in Cylindrical Pores: A Molecular Dynamics Study. *Food Biophysics* **2011**, *6*, 233-240.

24. Klameth, F.; Vogel, M., Slow Water Dynamics near a Glass Transition or a Solid Interface: A Common Rationale. *J. Phys. Chem. Lett.* **2015**, *6*, 4385-4389.

25. Liu, L.; Li, S.; Cao, Z.; Peng, Y.; Li, G.; Yan, T.; Gao, X.-P., Well-Ordered Structure at Ionic Liquid/Rutile (110) Interface. *J. Phys. Chem. C* **2007**, *111*, 12161-12164.

26. Bovio, S.; Podestà, A.; Lenardi, C.; Milani, P., Evidence of Extended Solidlike Layering in [Bmim][Ntf2] Ionic Liquid Thin Films at Room-Temperature. *J. Phys. Chem. B* **2009**, *113*, 6600-6603.

27. Hayes, R.; Borisenko, N.; Tam, M. K.; Howlett, P. C.; Endres, F.; Atkin, R., Double Layer Structure of Ionic Liquids at the Au(111) Electrode Interface: An Atomic Force Microscopy Investigation. *J. Phys. Chem. C* **2011**, *115*, 6855-6863.

28. Segura, J. J.; Elbourne, A.; Wanless, E. J.; Warr, G. G.; Voitchovsky, K.; Atkin, R., Adsorbed and near Surface Structure of Ionic Liquids at a Solid Interface. *Phys. Chem. Chem. Phys.* **2013**, *15*, 3320-3328.

29. Bovio, S.; Podestà, A.; Milani, P.; Ballone, P.; Pópolo, M. G. D., Nanometric Ionic-Liquid Films on Silica: A joint Experimental and Computational Study. *J. Phys. Condens. Matter* **2009**, *21*, 424118.

30. Piper, D. M., et al., Stable Silicon-Ionic Liquid Interface for Next-Generation Lithium-Ion Batteries. *Nat. Commun.* **2015**, *6*, 6230.

31. Perkin, S.; Salanne, M.; Madden, P.; Lynden-Bell, R., Is a Stern and Diffuse Layer Model Appropriate to Ionic Liquids at Surfaces? *Proc. Natl. Acad. Sci.* **2013**, *110*, E4121.

32. Kornyshev, A. A., Double-Layer in Ionic Liquids: Paradigm Change? *J. Phys. Chem. B* **2007**, *111*, 5545-5557.

33. Sha, M.; Wu, G.; Dou, Q.; Tang, Z.; Fang, H., Double-Layer Formation of [Bmim][Pf6] Ionic Liquid Triggered by Surface Negative Charge. *Langmuir* **2010**, *26*, 12667-12672.

34. Fitchett, B. D.; Conboy, J. C., Structure of the Room-Temperature Ionic Liquid/Sio2 Interface Studied by Sum-Frequency Vibrational Spectroscopy. *J. Phys. Chem. B* **2004**, *108*, 20255-20262.

35. Carmichael, A. J.; Hardacre, C.; Holbrey, J. D.; Nieuwenhuyzen, M.; Seddon, K. R., Molecular Layering and Local Order in Thin Films of 1-Alkyl-3-Methylimidazolium Ionic Liquids Using X-Ray Reflectivity. *Mol. Phys.* **2001**, *99*, 795-800.

36. Atkin, R.; Warr, G. G., Structure in Confined Room-Temperature Ionic Liquids. *J. Phys. Chem. C* **2007**, *111*, 5162-5168.

37. Liu, Y.; Zhang, Y.; Wu, G.; Hu, J., Coexistence of Liquid and Solid Phases of Bmim-Pf6 Ionic Liquid on Mica Surfaces at Room Temperature. *J. Am. Chem. Soc* **2006**, *128*, 7456-7457.

38. Maolin, S.; Fuchun, Z.; Guozhong, W.; Haiping, F.; Chunlei, W.; Shimou, C.; Yi, Z.; Jun, H., Ordering Layers of [Bmim][Pf6] Ionic Liquid on Graphite Surfaces: Molecular Dynamics Simulation. *J. Chem. Phys.* **2008**, *128*, 134504.

39. Sieffert, N.; Wipff, G., Ordering of Imidazolium-Based Ionic Liquids at the A-Quartz(001) Surface: A Molecular Dynamics Study. *J. Phys. Chem. C* **2008**, *112*, 19590-19603.

40. Federici Canova, F.; Mizukami, M.; Imamura, T.; Kurihara, K.; Shluger, A. L., Structural Stability and Polarisation of Ionic Liquid Films on Silica Surfaces. *Phys. Chem. Chem. Phys.* **2015**, *17*, 17661-17669.

41. Li, S.; Han, K. S.; Feng, G.; Hagaman, E. W.; Vlcek, L.; Cummings, P. T., Dynamic and Structural Properties of Room-Temperature Ionic Liquids near Silica and Carbon Surfaces. *Langmuir* **2013**, *29*, 9744-9749.

42. Shimizu, K.; Pensado, A.; Malfreyt, P.; Padua, A. A. H.; Canongia Lopes, J. N., 2d or Not 2d: Structural and Charge Ordering at the Solid-Liquid Interface of the 1-(2-Hydroxyethyl)-3-Methylimidazolium Tetrafluoroborate Ionic Liquid. *Faraday Discuss.* **2012**, *154*, 155-169.

43. Waechtler, M.; Sellin, M.; Stark, A.; Akcakayiran, D.; Findenegg, G.; Gruenberg, A.; Breitzke, H.; Buntkowsky, G., 2h and 19f Solid-State Nmr Studies of the Ionic Liquid [C2py][Bta]-D10 Confined in Mesoporous Silica Materials. *Phys. Chem. Chem. Phys.* **2010**, *12*, 11371-11379.

44. Rajput, N. N.; Monk, J.; Hung, F. R., Structure and Dynamics of an Ionic Liquid Confined inside a Charged Slit Graphitic Nanopore. *J. Phys. Chem. C* **2012**, *116*, 14504-14513.

45. Rivera-Rubero, S.; Baldelli, S., Influence of Water on the Surface of Hydrophilic and Hydrophobic Room-Temperature Ionic Liquids. *J. Am. Chem. Soc* **2004**, *126*, 11788-11789.

46. Huddleston, J. G.; Visser, A. E.; Reichert, W. M.; Willauer, H. D.; Broker, G. A.; Rogers, R. D., Characterization and Comparison of Hydrophilic and Hydrophobic Room Temperature Ionic Liquids Incorporating the Imidazolium Cation. *Green Chem.* **2001**, *3*, 156-164.

47. Cammarata, L.; Kazarian, S. G.; Salter, P. A.; Welton, T., Molecular States of Water in Room Temperature Ionic Liquids. *Phys. Chem. Chem. Phys.* **2001**, *3*, 5192-5200.

48. Berendsen, H. J. C.; van der Spoel, D.; van Drunen, R., Gromacs: A Message-Passing Parallel Molecular Dynamics Implementation. *Comput. Phys. Commun.* **1995**, *91*, 43-56.

49. Tokuda, H.; Hayamizu, K.; Ishii, K.; Susan, M. A. B. H.; Watanabe, M., Physicochemical Properties and Structures of Room Temperature Ionic Liquids. 1. Variation of Anionic Species. *J. Phys. Chem. B* **2004**, *108*, 16593-16600.

50. Gardas, R. L.; Freire, M. G.; Carvalho, P. J.; Marrucho, I. M.; Fonseca, I. M. A.; Ferreira, A. G. M.; Coutinho, J. A. P., High-Pressure Densities and Derived Thermodynamic Properties of Imidazolium-Based Ionic Liquids. *J. Chem. Eng. Data* **2007**, *52*, 80-88.

51. Bhargava, B. L.; Balasubramanian, S., Refined Potential Model for Atomistic Simulations of Ionic Liquid [Bmim][Pf6]. *J. Chem. Phys.* **2007**, *127*, 114510.

52. Canongia Lopes, J. N.; Deschamps, J.; Pádua, A. A. H., Modeling Ionic Liquids Using a Systematic All-Atom Force Field. *J. Phys. Chem. B* **2004**, *108*, 2038-2047.

53. Pádua, A. A. H.; Canongia Lopes, J. N., Molecular Force Field for Ionic Liquids. **version 2016/10/14**, *https://github.com/agiliopadua/ilff/blob/master/il.ff*

54. Sambasivarao, S. V.; Acevedo, O., Development of Opls-Aa Force Field Parameters for 68 Unique Ionic Liquids. *J. Chem. Theory Comput.* **2009**, *5*, 1038-1050.

55. Wieser, M. E.; Coplen, T. B., Atomic Weights of the Elements 2009 (Iupac Technical Report). *Pure Appl. Chem.* **2011**, *83*, 359-396.

56. Hanwell, M. D.; Curtis, D. E.; Lonie, D. C.; Vandermeersch, T.; Zurek, E.; Hutchison, G. R., Avogadro: An Advanced Semantic Chemical Editor, Visualization, and Analysis Platform. *J. Cheminform.* **2012**, *4*, 17.

57. Martínez, L.; Andrade, R.; Birgin, E. G.; Martínez, J. M., Packmol: A Package for Building Initial Configurations for Molecular Dynamics Simulations. *J. Comput. Chem.* **2009**, *30*, 2157-2164.

58. Emami, F. S.; Puddu, V.; Berry, R. J.; Varshney, V.; Patwardhan, S. V.; Perry, C. C.; Heinz, H., Force Field and a Surface Model Database for Silica to Simulate Interfacial Properties in Atomic Resolution.*Chem. Mater.* **2014**, *26*, 2647-2658.

59. Van Gunsteren, W. F.; Berendsen, H. J. C., A Leap-Frog Algorithm for Stochastic Dynamics. *Mol Simul.* **1988**, *1*, 173-185.

60. Darden, T.; York, D.; Pedersen, L., Particle Mesh Ewald: An N·Log(N) Method for Ewald Sums in Large Systems. *J. Chem. Phys.* **1993**, *98*, 10089-10092.

61. Essmann, U.; Perera, L.; Berkowitz, M. L.; Darden, T.; Lee, H.; Pedersen, L. G., A Smooth Particle Mesh Ewald Method. *J. Chem. Phys.* **1995**, *103*, 8577-8593.

62. Berendsen, H. J. C.; Postma, J. P. M.; Gunsteren, W. F. v.; DiNola, A.; Haak, J. R., Molecular Dynamics with Coupling to an External Bath. *J. Chem. Phys.* **1984**, *81*, 3684-3690.

63. Nosé, S., A Unified Formulation of the Constant Temperature Molecular Dynamics Methods. *J. Chem. Phys.* **1984**, *81*, 511-519.

64. Hoover, W. G., Canonical Dynamics: Equilibrium Phase-Space Distributions. *Phys. Rev. A* **1985**, *31*, 1695-1697.

65. Parrinello, M.; Rahman, A., Polymorphic Transitions in Single Crystals: A New Molecular Dynamics Method. *J. Appl. Phys.* **1981**, *52*, 7182-7190.

66. Laage, D.; Hynes, J. T., A Molecular Jump Mechanism of Water Reorientation. *Science* **2006**, *311*, 832-835.

67. Carosati, E.; Sciabola, S.; Cruciani, G., Hydrogen Bonding Interactions of Covalently Bonded Fluorine Atoms: From Crystallographic Data to a New Angular Function in the Grid Force Field. *J. Med. Chem.* **2004**, *47*, 5114-5125.

68. Howard, J. A. K.; Hoy, V. J.; O'Hagan, D.; Smith, G. T., How Good Is Fluorine as a Hydrogen Bond Acceptor? *Tetrahedron* **1996**, *52*, 12613-12622.

69. Sloutskin, E.; Ocko, B. M.; Tamam, L.; Kuzmenko, I.; Gog, T.; Deutsch, M., Surface Layering in Ionic Liquids: An X-Ray Reflectivity Study. *J. Am. Chem. Soc* **2005**, *127*, 7796-7804.

70. Pal, T.; Vogel, M., Role of Dynamic Heterogeneities in Ionic Liquids: Insights from All-Atom and Coarse-Grained Molecular Dynamics Simulation Studies. *Chem. Phys. Chem*, **2017**, *18*, 2233–2242.

71. Brammer, L.; Bruton, E. A.; Sherwood, P., Understanding the Behavior of Halogens as Hydrogen Bond Acceptors. *Crystal Growth & Design* **2001**, *1*, 277-290.

72. West, R.; Powell, D. L.; Whatley, L. S.; Lee, M. K. T.; von R. Schleyer, P., The Relative Strengths of Alkyl Halides as Proton Acceptor Groups in Hydrogen Bonding. *J. Am. Chem. Soc* **1962**, *84*, 3221-3222.

73. Lammert, H.; Kunow, M.; Heuer, A., Complete Identification of Alkali Sites in Ion Conducting Lithium Silicate Glasses: A Computer Study of Ion Dynamics. *Phys. Rev. Lett.* **2003**, *90*, 215901.

74. Vogel, M., Identification of Lithium Sites in a Model of $LiPO_3$ Glass: Effects of the Local Structure and Energy Landscape on Ionic Jump Dynamics. *Phys. Rev. B* **2004**, *70*, 094302.

75. Richard, H. H., Water's Dual Nature and Its Continuously Changing Hydrogen Bonds. *J. Phys. Condens. Matter* **2016**, *28*, 384001.

76. Skarmoutsos, I.; Welton, T.; Hunt, P. A., The Importance of Timescale for Hydrogen Bonding in Imidazolium Chloride Ionic Liquids. *Phys. Chem. Chem. Phys.* **2014**, *16*, 3675-3685.

77. Aliaga, C.; Santos, C. S.; Baldelli, S., Surface Chemistry of Room-Temperature Ionic Liquids. *Phys. Chem. Chem. Phys.* **2007**, *9*, 3683-3700.

**Figure Captions**

**Fig. 1:** Schematic representations of the 1-butyl-3-methylimidazolium [Bmim] cation and the hexafluorophosphate [PF$_6$] and tetrafluoroborate [BF$_4$] anions.

**Fig. 2:** Number density profiles $n_z$ of IL cations (red) and anions (blue) across the silica pore normalized by the corresponding bulk values $n_0$. The main panel shows data for [Bmim][PF$_6$] across the whole slit at 300 K. The grey shaded area marks the location of the amorphous silica slab, which is reproduced on the right hand side due to periodic boundary conditions. The inserted panel focuses on the interfacial region and compares results for [Bmim][PF$_6$] (solid lines) and [Bmim][BF$_4$] (dashed lines) at 300 K.

**Fig. 3:** Probability distribution of cos $\theta_{butyl}$ for [Bmim] cations in the first layer at the silica surface. As sketched, $\theta_{butyl}$ species the orientation of the butyl tail of [Bmim] with respect to the $z$ axis, i.e., the normal vector of the silica surface. Results for [Bmim][BF$_4$] and [Bmim][PF$_6$] at 300 K are shown.

**Fig. 4:** Probability distribution of cos $\theta_{ring}$ for [Bmim] cations in the first layer at the silica surface. As sketched, $\theta_{ring}$ species the orientation of the normal vector of the imidazolium ring with respect to the $z$ axis, i.e., the normal vector of the silica surface. Results for [Bmim][BF$_4$] and [Bmim][PF$_6$] at 300 K are shown.

**Fig. 5:** Typical cation arrangements at the silica surface for (top) [Bmim][PF$_6$] and (bottom) [Bmim][BF$_4$] at 300 K. All cations located in the first layer at a given time frame are shown together with the contour of the amorphous silica surface. Yellow regions indicate the positions of the silicon atoms, whereas red and white regions are populated by the oxygen and hydrogen atoms of the hydroxyl groups.

**Fig. 6:** Mean square displacement of (a) all cations and (b) all anions of [Bmim][PF$_6$] and [Bmim][BF$_4$] in the silica pore at 300 K.

**Fig. 7:** Temperature dependence of inverse self-diffusivities ($D^{-1}$) of cations (red) and anions (blue) in [Bmim][PF$_6$] (squares) and [Bmim][BF$_4$] (triangles) systems. Results for the confined (open symbols) and bulk (solid symbols) ILs are compared. The data for confined [Bmim][BF$_4$] are fitted to the Arrhenius law, the other data are interpolated with the VFT equation. The fits are shown as lines.

**Fig. 8**: Spatially resolved incoherent scattering functions $S(q,t)$ characterizing (a) cation dynamics and (b) anion dynamics in [Bmim][PF$_6$] at 300 K. The analysis uses a scattering vector of $q$ = 14.50 nm$^{-1}$ and distinguishes between ions in different layers at the silica surface. For comparison, we include results obtained from the average over the whole pore volume and data of the bulk system at 300 K, as obtained for a similar scattering vector of $q$ = 15.70 nm$^{-1}$.

**Fig. 9:** Temperature-dependent structural relaxation times obtained from the spatially resolved incoherent intermediate scattering functions in Fig. 8. Results for [PF$_6$] (squares) and [BF$_4$] (triangles) anions in the first (solid symbols) and third (striped symbols) layers, respectively, are compared with that obtained from an average over the whole pore volume (open symbols).

**Fig. 10:** 2D representation (*xy*-plane) of spatial regions in the first layer at the silica wall, which have a high probability density to host (top) P atoms of [PF$_6$] and (bottom) B atoms of [BF$_4$] at 350 K. Various adsorption sites of the anions at the surface are clearly distinguishable and are given different colors and sequential numbers. Triangles indicate the centers of the sites. Circles mark the time-averaged positions of the oxygen atoms of the silanol groups. Silanol groups that can form a hydrogen bond with anions residing at a particular site have the same color as this site.

**Fig. 11:** Characteristic properties of anionic adsorption sites at 350 K: (a) site volume $V_s$, (b) time-averaged number $\langle n_{anion} \rangle$ of anions occupying the site, and (c) number of silanol groups $n_{SiOH}$ associated with the site. The site numbers are defined in Fig. 10. The solid blue and red

lines show results for [PF6] and [BF4] sites, respectively. The corresponding dashed lines indicate the respective averages over all sites identified for a given anionic species.

**Fig. 12:** Cumulative distribution function of the time-averaged number $\langle n_{HB} \rangle$ of hydrogen bonds formed by each of the silanol groups with an anion of the given species. Results for [BF4] and [PF6] are shown as dashed and solid lines, respectively. The vertical line marks the fully bonded state $\langle n_{HB} \rangle = 1$.

**Fig. 13:** Examples of configurations of (top) [BF4] and (bottom) [PF6] at adsorption sites at 350 K. The red and white spheres mark the oxygen and hydrogen atoms of the silanol groups. The dotted lines mark the hydrogen-bonds, including furcated bonds. The numbers indicate the distances in angstrom between the fluorine and the silanol hydrogen atoms.

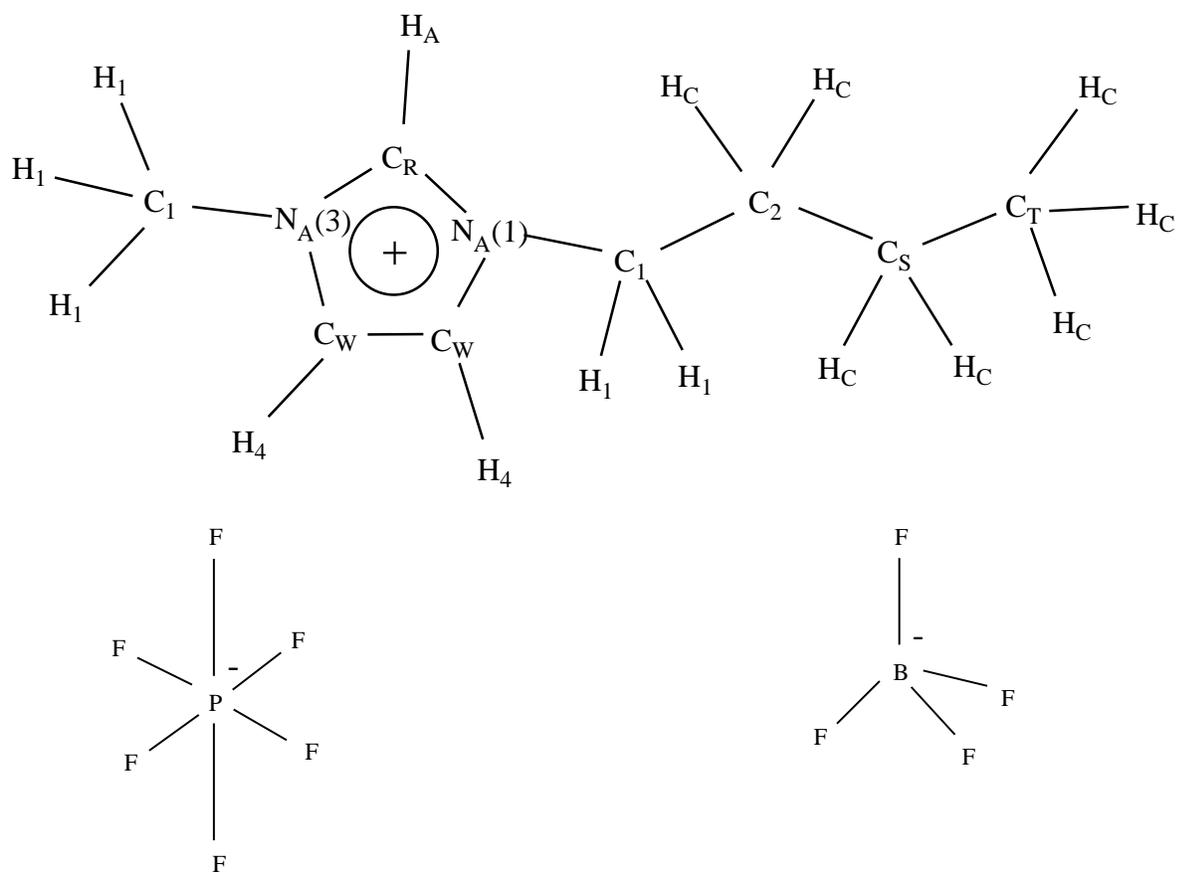

Fig. 1/ Pal, Beck, Lessnich & Vogel

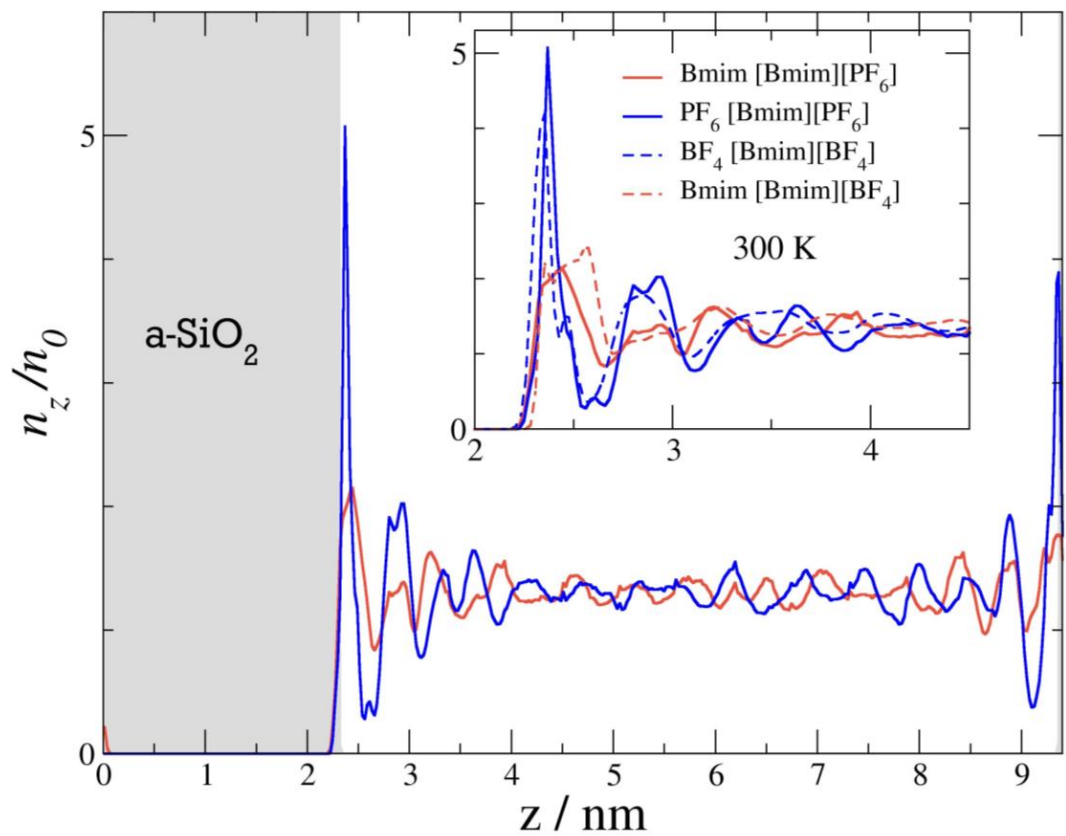

Fig. 2/ Pal, Beck, Lessnich & Vogel

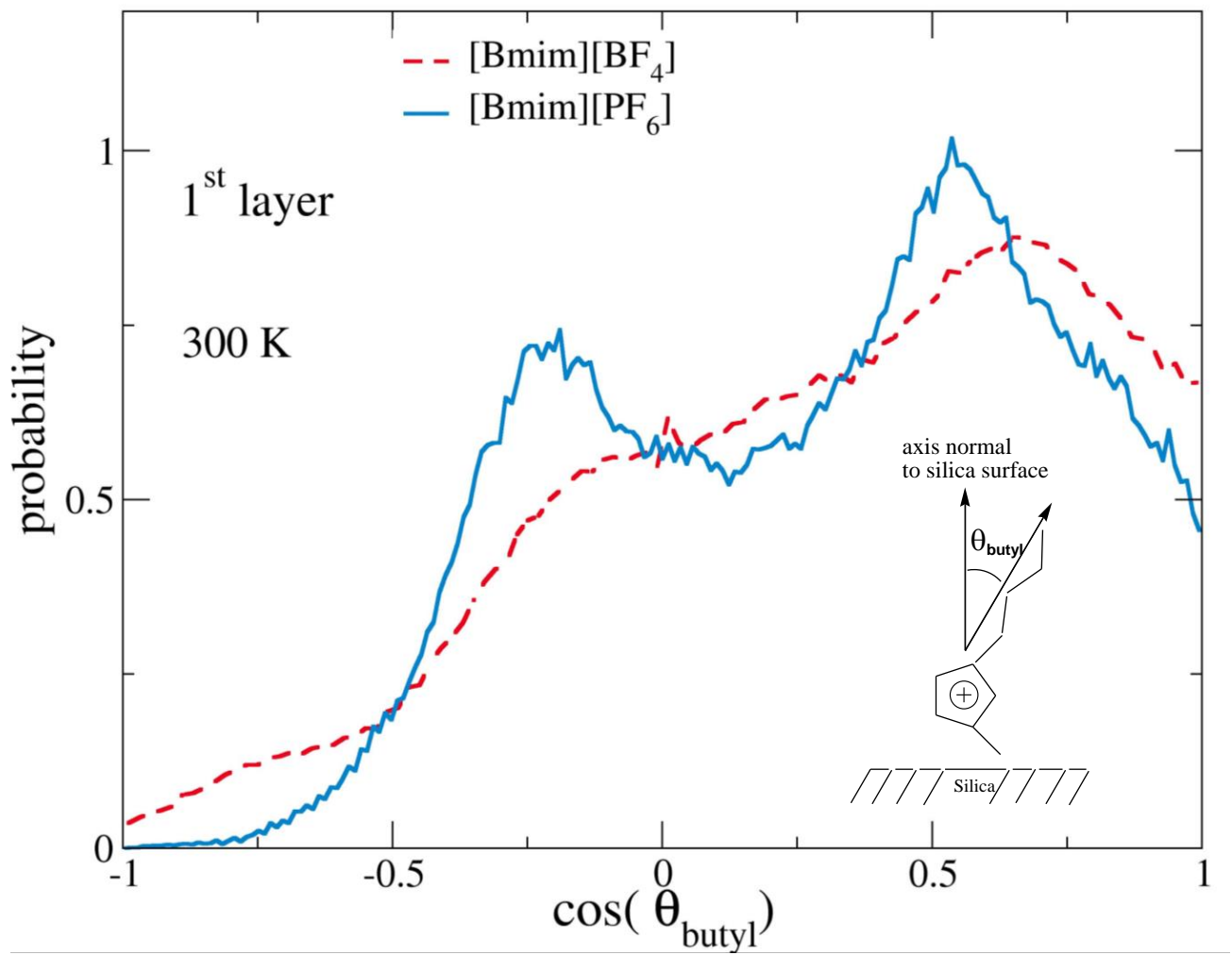

Fig. 3/ Pal, Beck, Lessnich & Vogel

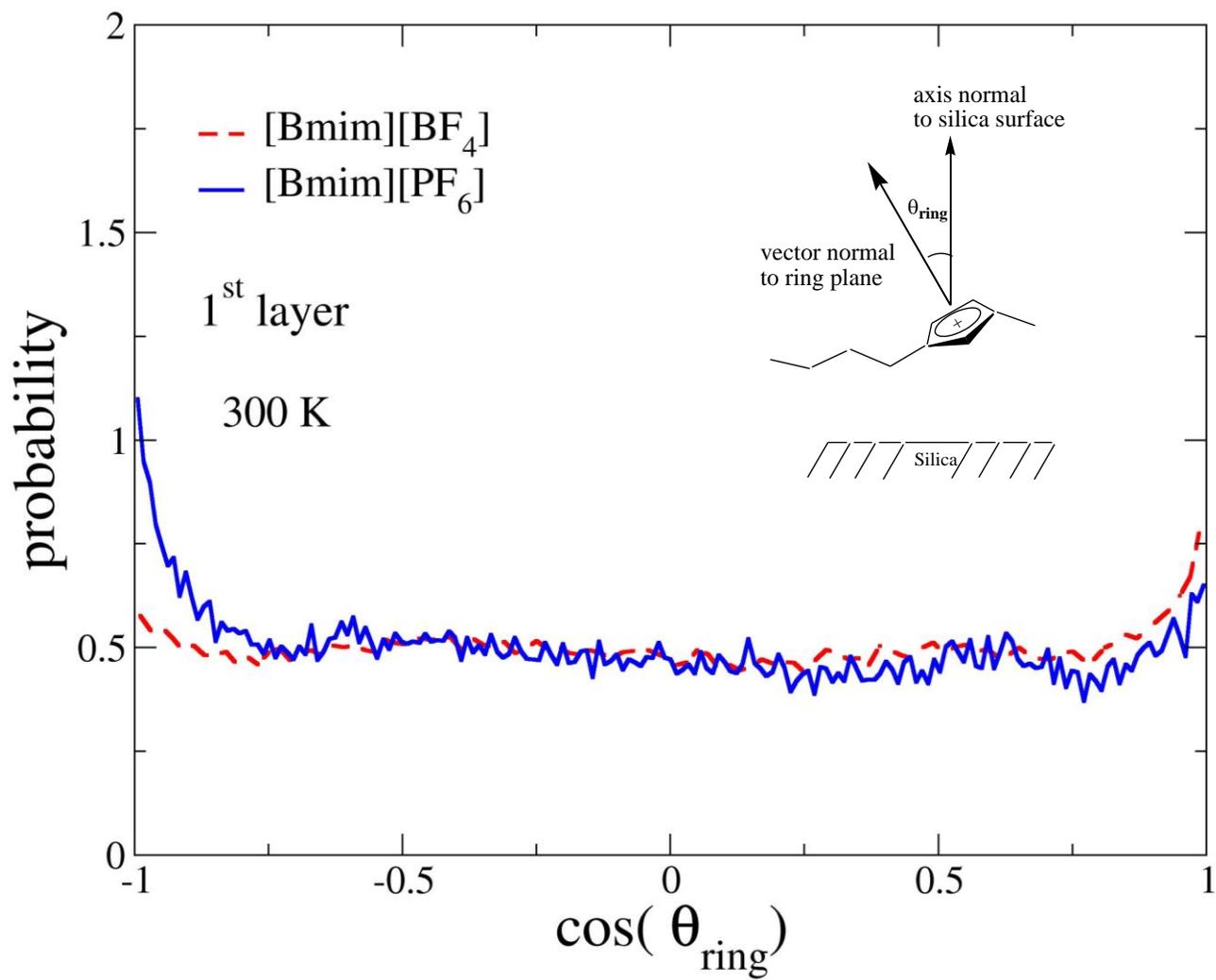

Fig. 4/ Pal, Beck, Lessnich & Vogel

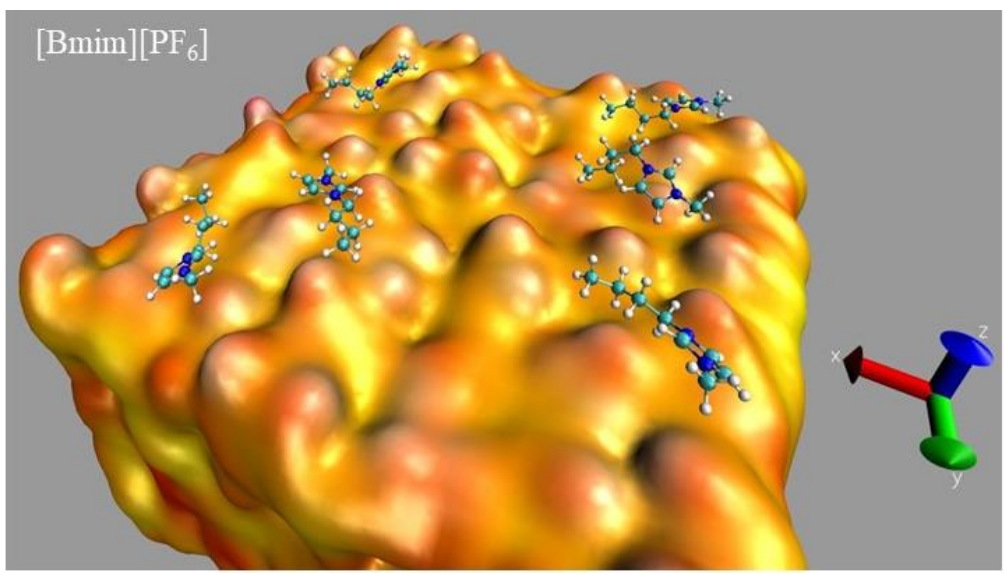

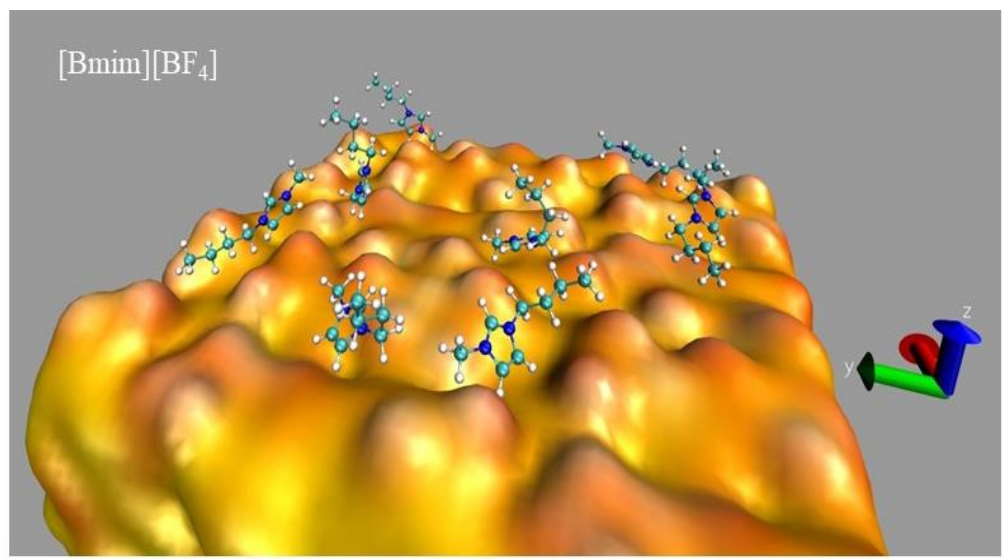

Fig. 5/ Pal, Beck, Lessnich & Vogel

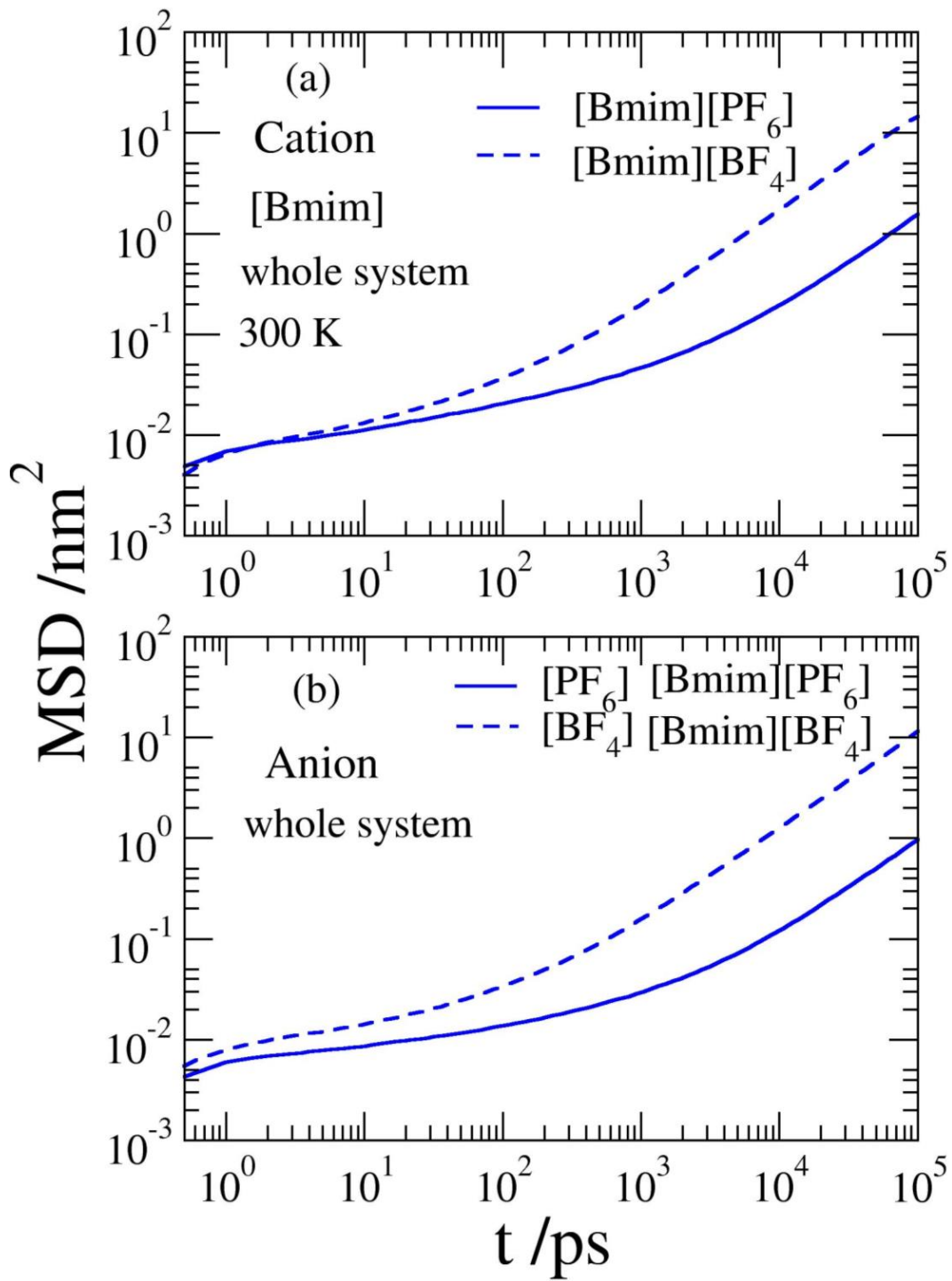

Fig. 6/ Pal, Beck, Lessnich & Vogel

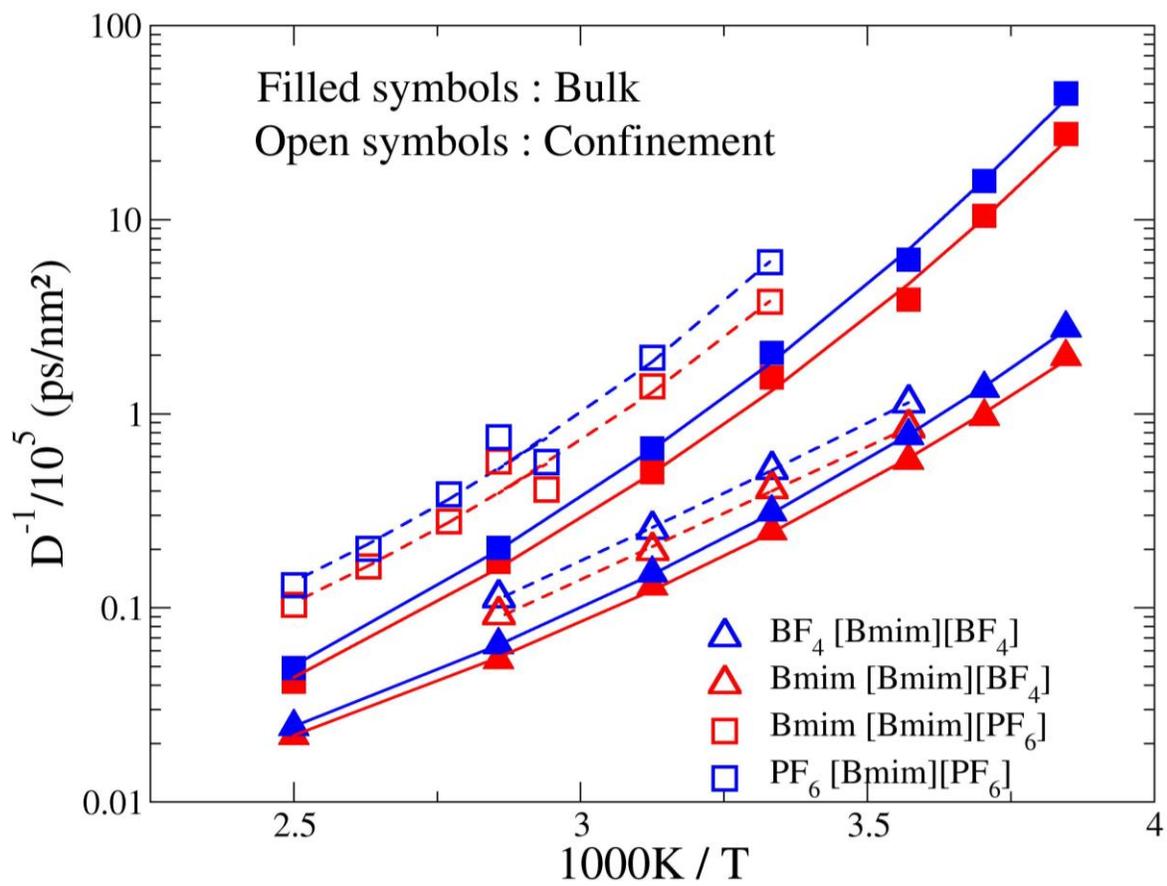

Fig. 7/ Pal, Beck, Lessnich & Vogel

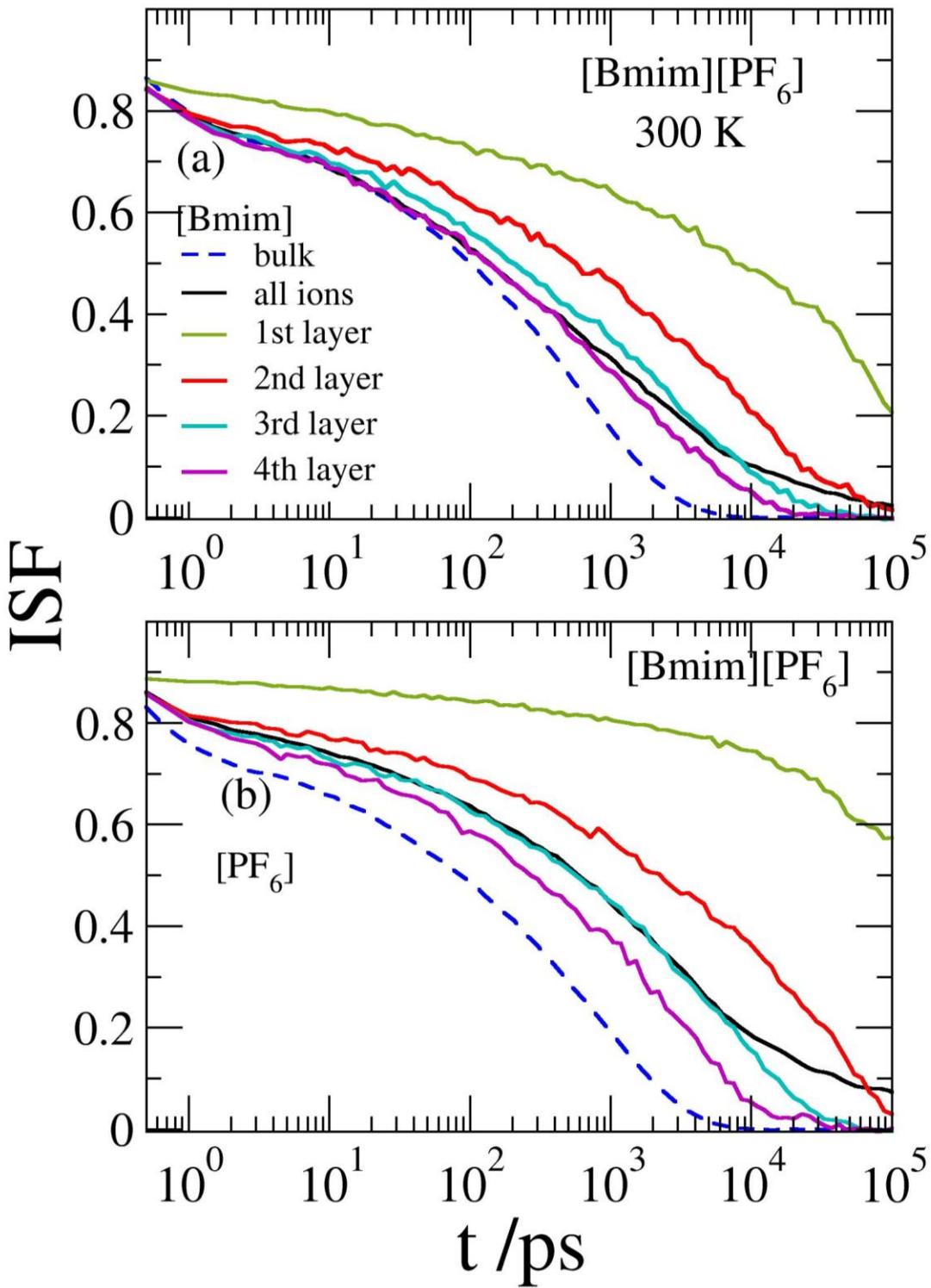

Fig. 8/ Pal, Beck, Lessnich & Vogel

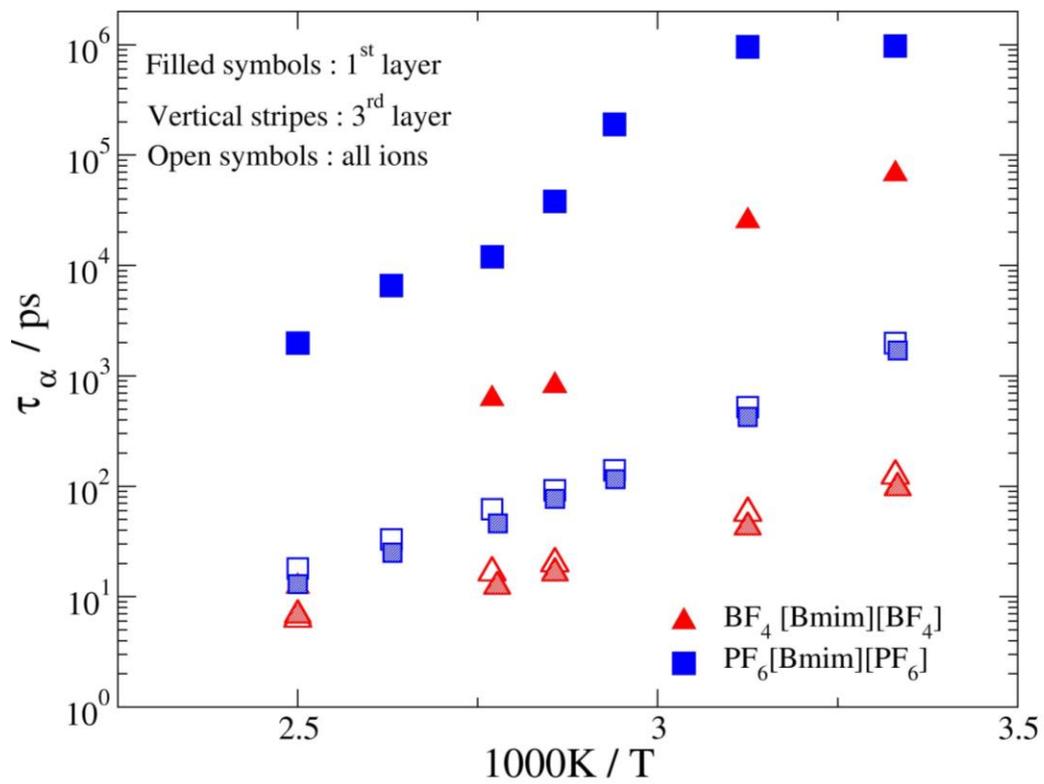

Fig. 9/ Pal, Beck, Lessnich & Vogel

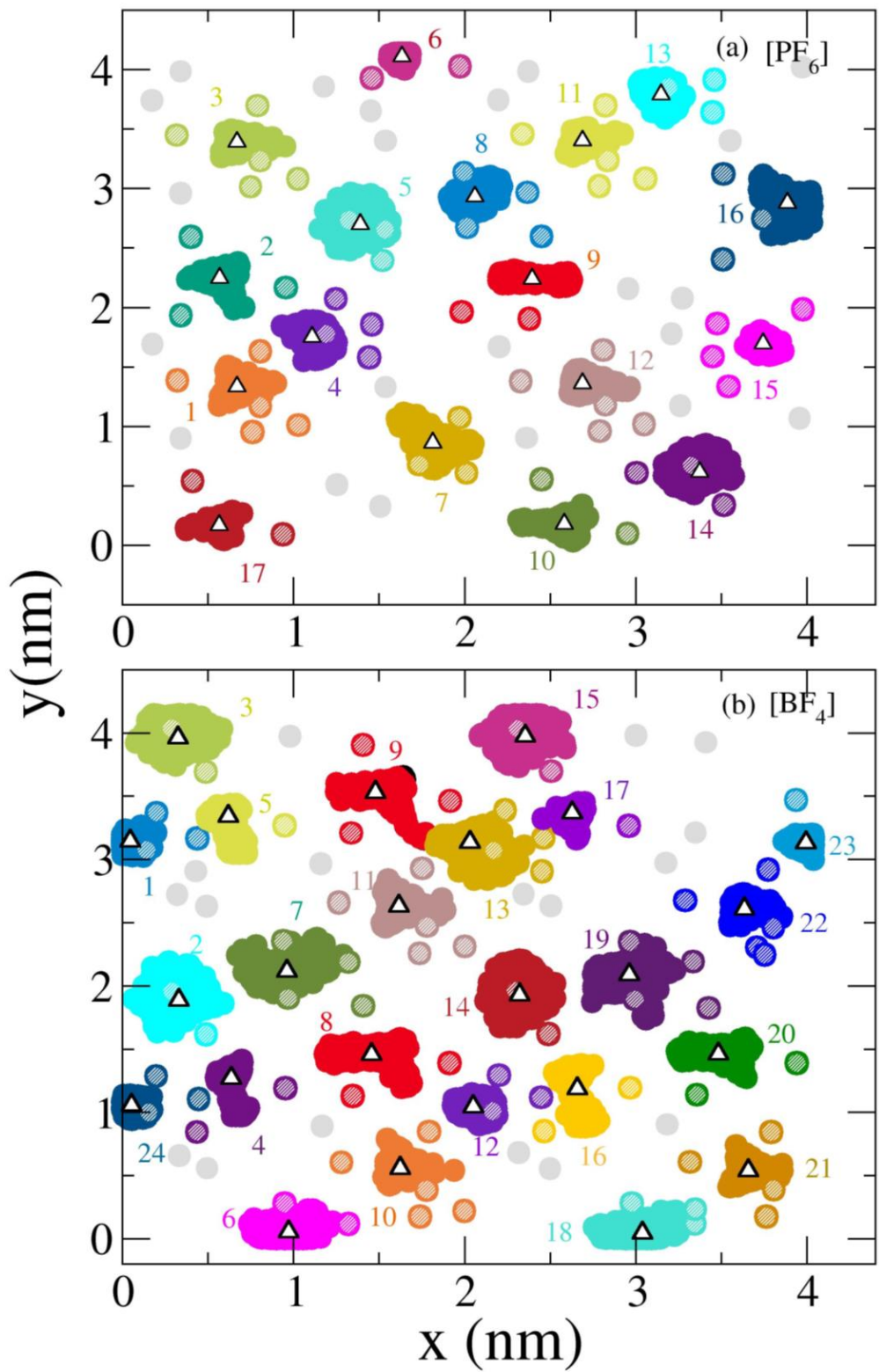

Fig. 10/ Pal, Beck, Lessnich & Vogel

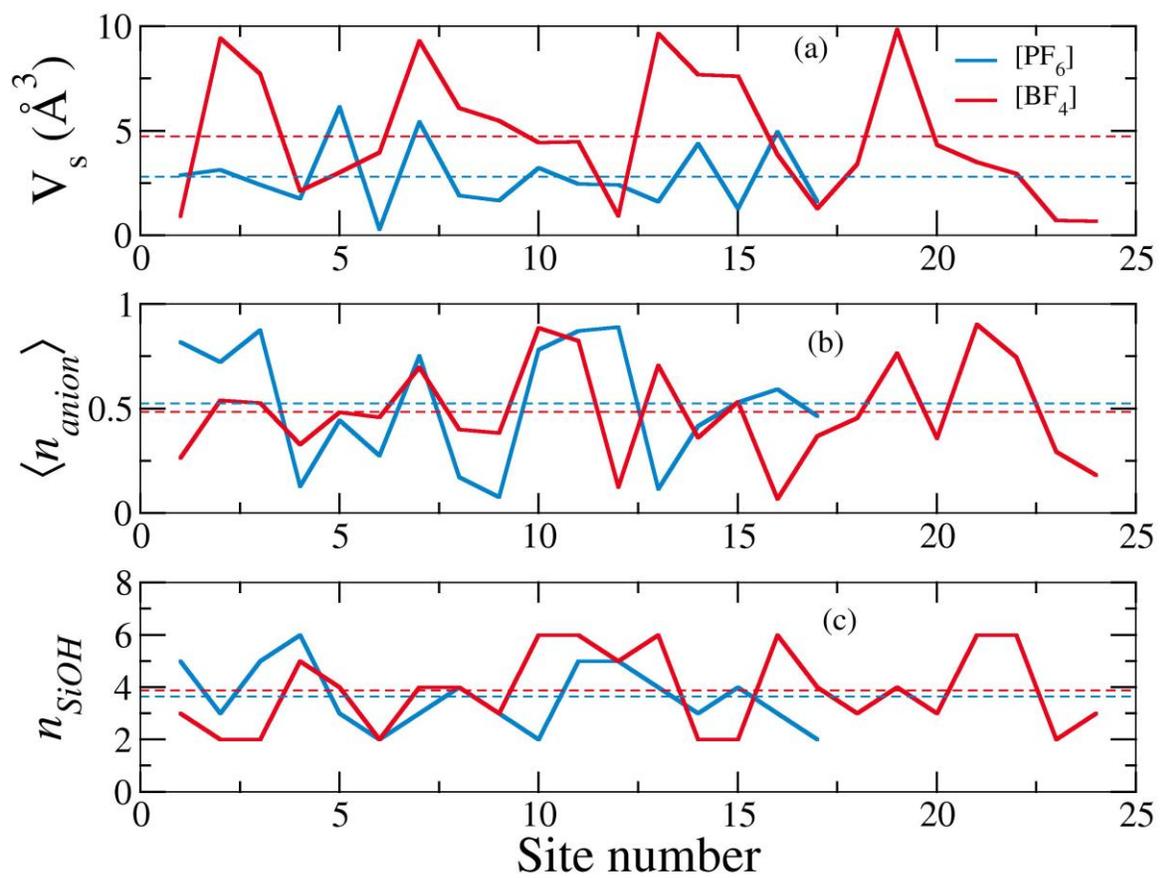

Fig. 11/ Pal, Beck, Lessnich & Vogel

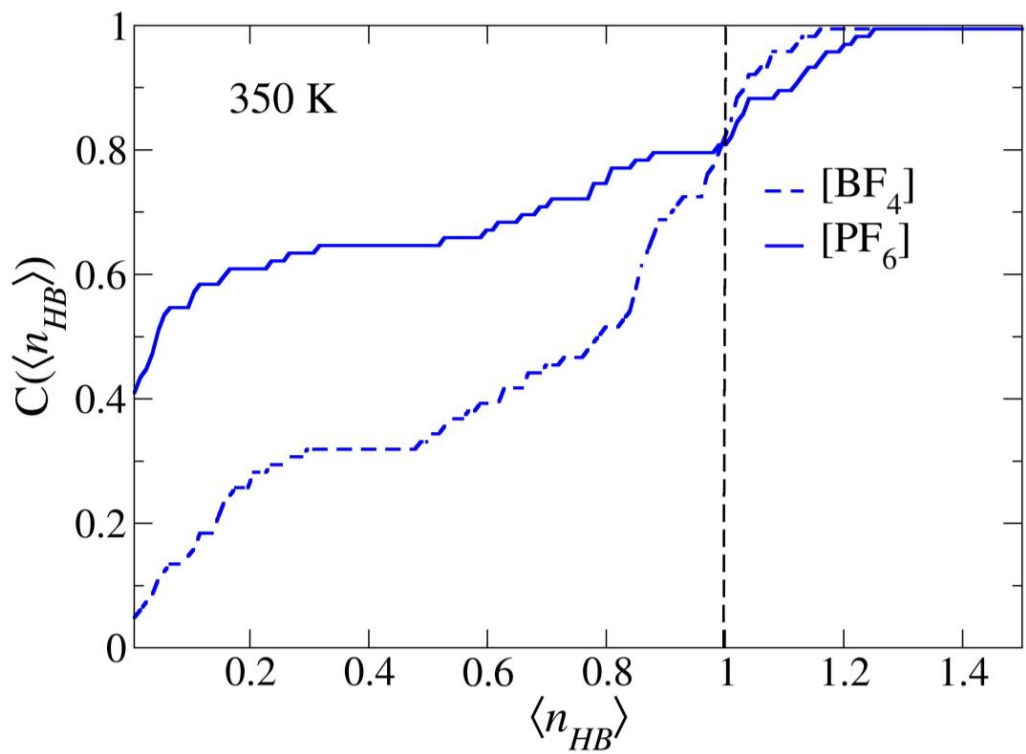

Fig. 12/ Pal, Beck, Lessnich & Vogel

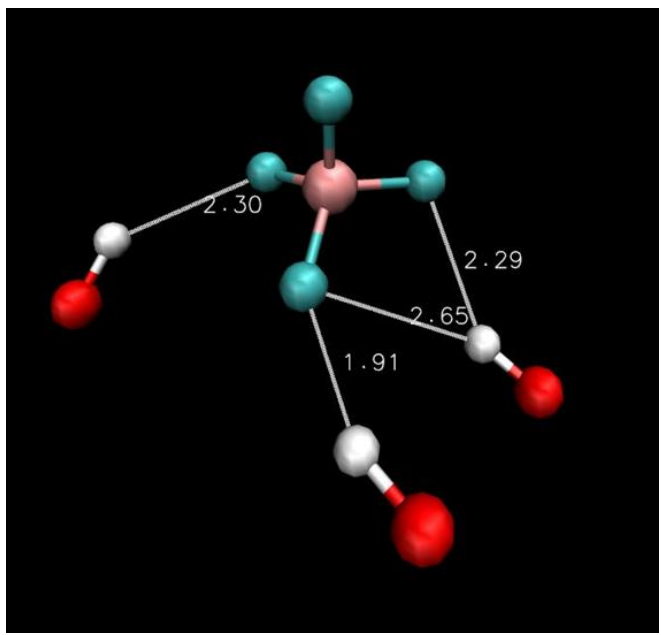
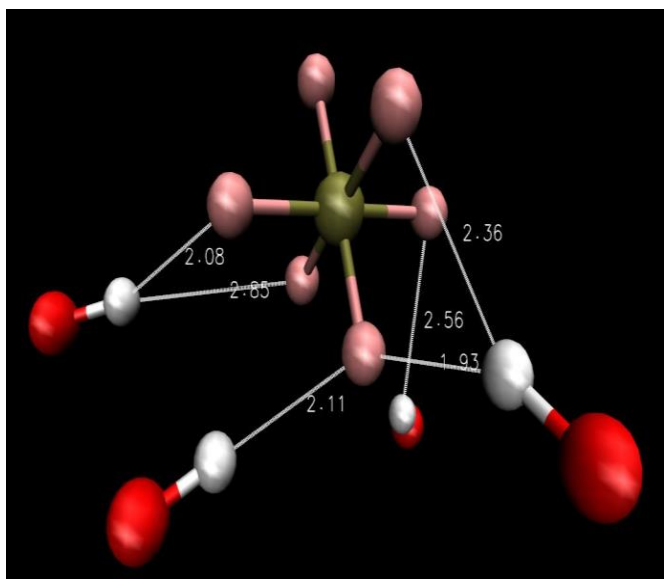

Fig. 13/ Pal, Beck, Lessnich & Vogel

Table-of-Content (TOC) Graphic for the Article entitled, "Effects of Silica Surfaces on the Structure and Dynamics of Room Temperature Ionic Liquids: A Molecular Dynamics Simulation Study" by T. Pal, C. Beck, D. Lessnich & M. Vogel

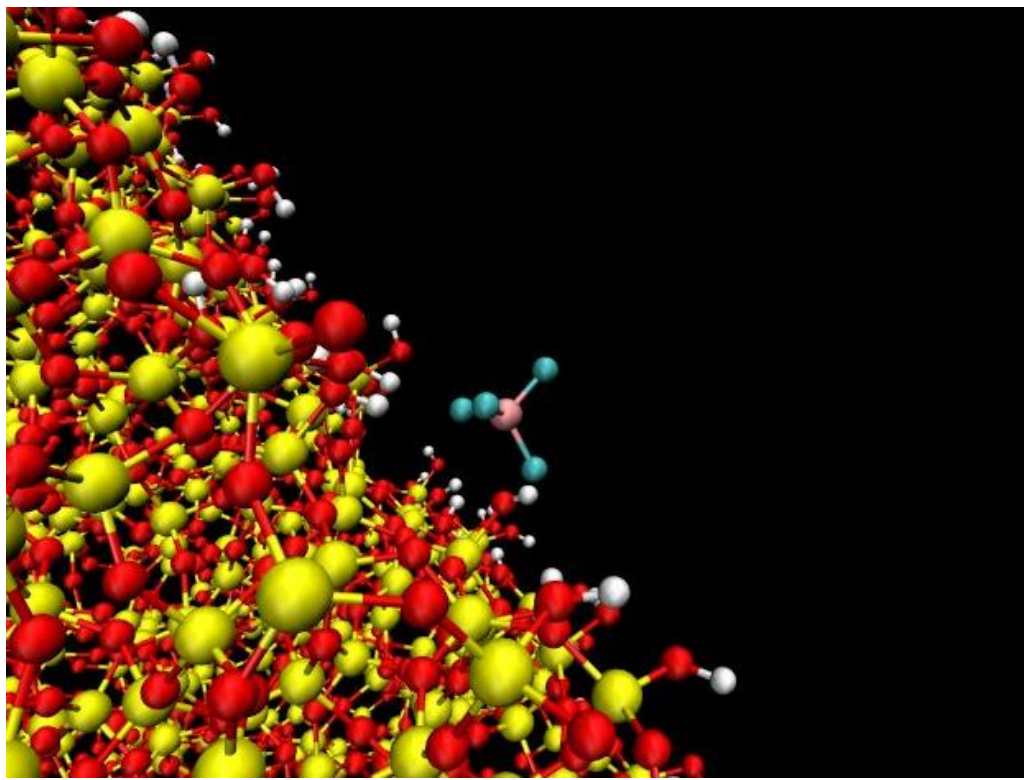